\documentclass[twocolumn,showpacs,amsmath,amssymb,prl]{revtex4}
\usepackage{dcolumn}
\usepackage{epsfig}
\usepackage[latin1]{inputenc} 
\usepackage{bm}

\begin{document}

%
%
%
\title{Linking rigidity transitions with enthalpic changes at the glass transition and the fragility of glass-forming liquids}
\author{Matthieu Micoulaut}
\affiliation{Laboratoire de Physique Théorique de la Matière Condensée,
UPMC - Université Paris 6,  Boite 121\\ 4, Place Jussieu, 75252
Paris Cedex 05, France\\}

\date{\today}
\begin{abstract}
A low temperature Monte Carlo dynamics of a Keating like oscillator model is used to study the relationship between the nature of glasses from the viewpoint of rigidity, and the strong-fragile behaviour of glass-forming liquids. The model shows that a Phillips optimal glass formation with minimal enthalpic changes is obtained under a cooling/annealing cycle when the system is optimally constrained by the harmonic interactions, i.e. when it is isostatically rigid. For these peculiar systems, the computed fragility shows also a minimum, which demonstrates that isostatically rigid glasses are strong (Arrhenius-like) glass-forming liquids. Experiments on chalcogenide and oxide glass-forming liquids are discussed under this new perspective and confirm the theoretical prediction for chalcogenide network glasses.
\end{abstract}
\maketitle
The question of the liquid to glass transition and the nature of its characteristic temperature $T_g$ have received a huge amount of interest in the recent years with a special emphasis on the dynamic properties of the glass-forming liquid \cite{Sciortino1}-\cite{Tg3}. Experimentally, the slowing down of the dynamics is mostly tracked from viscosity (or structural relaxation time $\tau_\alpha$) measurements. The behaviour of these quantities with inverse temperature not always displays an Arrhenius-like (or strong) behaviour. When properly rescaled with $1/T_g$ in a semi-log plot, the viscosity or relaxation time can indeed display a variety of different behaviours usually quantified by a fragility index $M$ \cite{Angellb} which characterizes the steepness of the slope of the relaxation time near the glass transition:
\begin{eqnarray}
\label{M}
M={\frac {d\ log_{10}\tau_\alpha}{d\ T/T_g}}\biggl]_{T=T_g}
\end{eqnarray}
$M$ ranges typically between $16$ for the strong (Arrhenius behaving) silica liquid, whereas $M$ is about $340$ for the fragile ortho-terphenyl which changes its viscosity by 10 orders of magnitude over only 50 K temperature change. 
\par
There have been various efforts to connect the liquid fragility to some easily measurable quantities in the glassy state such as compressibility \cite{Buchenau} or Poisson ratio \cite{Sokolov}, or to the out-off equilibrium behaviour \cite{Scopigno}. At a somewhat more microscopic level, combined effects of the structure and the local mechanical behaviour arising from the interaction potential has to act on the dynamical and calorimetric behaviour at the glass transition and thus on the fragility. From a theoretical viewpoint, one may therefore wonder how the effect of the potential strength in the glassy state on the relaxing behaviour at higher temperatures can be determined from simple but insightful models. 
\par
The present Letter attempts to address this basic issue by following the Monte Carlo dynamics of a harmonic Keating-like oscillator model that mimicts the elastic features (flexible, isostatic, stressed rigid) of the glass. The solution of the model shows that isostatic (i.e. optimally constrained or intermediate) glasses exhibit a minimum in the energy change during a cooling-annealing cycle through the glass transition, independently of the applied cooling rate, a result that matches exactly the Phillips optimal glass condition \cite{JCP79}. Furthermore, the present minimum is correlated with the minimum obtained in the fragility index. These findings are successfully confronted with experimental data on chalcogenide network glasses. One can thus conclude that isostatic glasses are strong glass-forming liquids, a result that opens new perspectives for the description of glass-forming liquids from the viewpoint of the mechanical behaviour of the glass. And since rigidity can be also tuned in soft solids and colloids \cite{Wyart}, it may provide a general clue for an improved understanding of the dynamic properties leading to a strong viscosity behaviour.
\par
The elastic nature of network glasses with changing connectivity (or mechanical constraints $n_c$ derived from Lagrange-Maxwell counting \cite{Maxwell}) can be modelled in the framework of rigidity theory \cite{Thorpe83}, \cite{Thorpe85} using a Keating potential that represents a semi-empirical description of covalent bond-stretching (BS) and bond-bending (BB) forces. From this description, it appears that the number of zero frequency (floppy) modes $f$ behaves as $3-n_c$. The glass composition at which one has the vanishing of $f$ corresponds to the Maxwell-Lagrange isostatic condition $n_c=3$. 
We consider a network of $N$ atoms having two types of harmonic oscillators with respective density $f$ and $n_c$. The first has a typical frequency $\omega$ associated with the harmonic motion due to the floppy modes, i.e. the modes that allow a local distorsion of the network with a low cost in energy. The second has a typical frequency $\Omega$ and represents the Keating potential containing the constraints imposed by BB and BS forces \cite{Wooten,Djork}. A similar approach has been recently used to describe the effect of rigidity on the glass transition temperature \cite{Naumis}. The interaction potential V is then given by: 
\begin{eqnarray}
\label{V}
V=e+E={\frac {1}{2}}\sum_i^Nf\omega^2x_i^2+n_c\Omega^2X_i^2
\end{eqnarray}
where $\omega$ and $\Omega$ represent respectively the floppy mode frequency and a typical vibrational mode related to BS or BB interactions. 
An inelastic neutron study \cite{Kamitakahara} of the ternary network glass Ge-As-Se provides information about the order of magnitude of the floppy mode, and the typical BS and BB vibrational frequencies (energies). The floppy mode energy is about $4~meV$ while indentified bond-stretching and bond-bending vibrations have a respective energy of $31$ and $19~meV$, i.e. about six times more than for the floppy mode energy. Thus one has $\Omega\simeq 6\omega$ and this ratio will be used for the forthcoming numerical applications.
\par
{\em Glass transition:} We now build on the approach developed by Ritort and co-workers \cite{Ritort1} \cite{Ritort2}, i.e. an energetical move $\Delta V$ is realized on the oscillators. It is applied according to the Metropolis algorithm, i.e. accepted with probability $1$ if the energy decreases, otherwise with a probability $\exp(-\beta\Delta V)$.
The position of the Keating and the floppy mode oscillators are
simultaneously shifted by $X_i+R_i/\sqrt{N}$ and $x_i+r_i/\sqrt{N}$ where $R_i$ and $r_i$ are random variables with a
Gaussian distribution having zero mean and respective variance $\Delta^2$ and $\delta^2$.  
\par
In comparing the nature of both oscillators, one obviously has $\delta\gg\Delta$ as motion around the position $x_i$ is facilitated by floppy modes, whereas the amplitude $X_i$ of the 'Keating' oscillators should be restricted to small vibrations, typically a fraction of the interatomic bond distance \cite{Mousseau}. 
The transition probability for the change in energy $\Delta V$ is :
\begin{eqnarray}
\label{dV}
P(\Delta V)=\int_{-\infty}^{\infty}
\biggl(\prod_i{\frac {dR_i}{\sqrt{2\pi\Delta^2}}}\exp{(-R_i^2/2\Delta^2)}\biggr)\\ \nonumber
\biggl(\prod_i{\frac {dr_i}{\sqrt{2\pi\delta^2}}}\exp{(-r_i^2/2\delta^2)}\biggr)
\\ \nonumber
\delta\biggl[\Delta V-n_c\Omega^2\sum_i\biggl({\frac {X_iR_i}{\sqrt{N}}}+{\frac {R_i^2}{N}}\biggr)-\\ \nonumber
f\omega^2\sum_i\biggl({\frac {x_ir_i}{\sqrt{N}}}+{\frac {r_i^2}{N}}\biggr)\biggr]
\end{eqnarray}
Using the Fourier transform representation of the delta function, equ. (\ref{dV}) 
can be expressed as a Gaussian in the limit $N\rightarrow\infty$ as:
\begin{eqnarray}
\label{dV1}
P(\Delta V)=\sqrt{\frac {1}{4\pi Q(t)}}\exp\biggl(-{\frac {(\Delta V-V_0)^2}{4Q(t)}}\biggr)
\end{eqnarray}
with $V_0={\frac {1}{2}}[f\omega^2\delta^2+n_c\Omega^2\Delta^2]$ and:
\begin{eqnarray}
\label{variance}
Q(t)=n_c\Omega^2\Delta^2<E(t)>+f\omega^2\delta^2<e(t)>
\end{eqnarray}
where the mean position of the oscillators have been taken as zero and the averages performed over different dynamical 
histories but with same initial condition for the ensemble. 
From the Metropolis rule, one can then write the equation for the evolution of the energy:
\begin{eqnarray}
\label{EvolE1}
{\frac {\partial V}{\partial t}}=\int_{-\infty}^0xP(x)dx+\int_0^{\infty}xP(x)e^{-\beta x}dx
\end{eqnarray}
which, applied to the probability distribution (\ref{dV1}) leads to:
\begin{eqnarray}
\label{EvolE2}
{\frac {\partial V}{\partial t}}={\frac {V_0}{2}}\bigg[\biggl(1-{\frac {2\beta Q(t)}{V_0}}\biggr)g(t)+erfc\biggl({\frac {V_0}{2\sqrt{Q(t)}}}\biggr)\biggr]
\end{eqnarray}
with $erfc$ the error function, and
\begin{eqnarray}
g(t)&=&\exp\biggl(-\beta V_0+\beta^2Q(t)\biggr)erfc\biggl[{\frac {2\beta Q(t)-V_0}{2\sqrt{Q(t)}}}\biggr]
\end{eqnarray}
\begin{figure}
\begin{center}
\epsfig{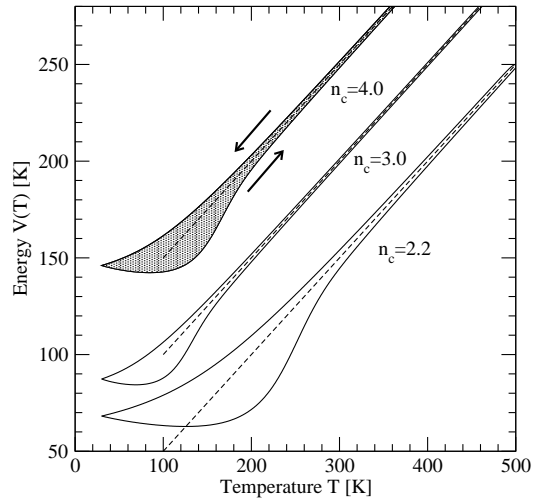}
\end{center}
\caption{\label{energy} Top panel: Energy $V(T)$ of the system (solution of equ. \ref{EvolE2}) under cooling and annealing ($q=\pm 1~K/s$) for three selected systems: $n_c=2.2$ (flexible), $n_c=3.0$ (isostatically rigid), $n_c=4.0$ (stressed rigid). The two last curves have been shifted upwards for a clearer presentation. Broken lines represent the equilibrium state $T/2$. The shaded areas (e.g. on $n_c=4$) serve to quantify the enthalpic changes ($H$).}
\end{figure}
Equation is not closed, because it depends on $Q(t)$ through the time dependence (equ. (\ref{variance})) of the averages of the floppy mode or constraint energies $e(t)$ and $E(t)$. For simplicity, we solve equation (\ref{EvolE1}) in the low temperature-long time adiabatic approximation \cite{Ritort3} where the derivative of the energy V iz zero. One then has:
\begin{eqnarray}
Q(t)={\frac {T}{2}}(2V_0)\simeq 2V_0V(t)
\end{eqnarray}
First, one can solve equation (\ref{EvolE2}) to obtain the behaviour of the energy $V(T)$. 
In order to highlight the effect of the number of constraints $n_c$ on $V(T)$, we work at fixed cooling/heating rate $q=\dot T=\pm1~K.s^{-1}$ and use $T_0=500~K$, $\delta=10\Delta$, $\Omega=0.3$.
Figure \ref{energy} represents the evolution of $V(T)$ for three different values of $n_c$. First, one notes that at high temperature, the energy of the system is equal to $T/2$ which is also the equilibrium solution of equ. (\ref{EvolE2}), in agreement with the equipartition theorem. Glassy behaviour (i.e. deviation from the T/2 line) onsets at lower temperature, defining a glass transition region around $150~K$. Annealing from a low temperature end point shows the typical hysteresis behaviour that is usually manifested in experiment by an enthalpic overshoot in the heat capacity \cite{Bool}. From Figure \ref{energy}, one sees that the enthalpic change between the cooling and heating curves (quantified by the area) depend on the number of mechanical constraints. Furthermore, these changes are minimized for a system that is nearly isostatically rigid ($n_c=3$), i.e. when the trial moves can be only realized on the oscillator with the highest frequency $\Omega$. Additional stiffening (i.e. increase of $n_c$) of the system for $n_c>3$ leads to a global increase of $V_0$ and thus to an increase of the area.
\par
\begin{figure}
\begin{center}
\epsfig{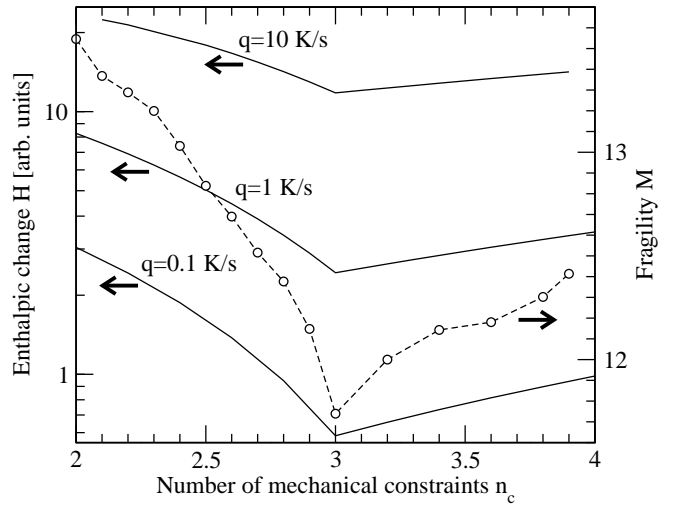}
\end{center}
\caption{\label{area} Enthalpic changes at the glass transition as a function of the number of mechanical constraints $n_c$ for different cycles at fixed cooling/heating rates $q=0.1~K/s$, $q=1~K/S$ and $q=10~K/s$. Right axis: fragility of the system.}
\end{figure}

 
The area $H$ defined by the difference between the cooling and heating curves can be tracked with the number of constraints $n_c$ and the corresponding behaviour is displayed in Figure \ref{area}. It shows that for a fixed cooling/heating rate $q$, certain glass transitions occur with minimal enthalpic changes. The model is therefore able to reproduce the Phillips phenomonelogy of ideal glass formation when a glass is optimally constraint, i.e. isostatic \cite{JCP79,Thorpe83}. This is experimentally observed on a characteristic enthalpy $\Delta H_{nr}$ (see also Fig. \ref{area}) extracted from complex heat flow measurements at the glass transition \cite{Bool1}.
\par
{\em Dynamics}: The study of the dynamics of the system can be achieved from the linearization of equation (\ref{EvolE2}) in the vicinity of the equilibrium value of $V(T)$ that leads to a typical relaxation time:
\begin{eqnarray}
\tau={\frac {1}{2}}\sqrt{\pi T^3V_0^3}\exp{V_0/4T}
\end{eqnarray}

\begin{figure}
\begin{center}
\epsfig{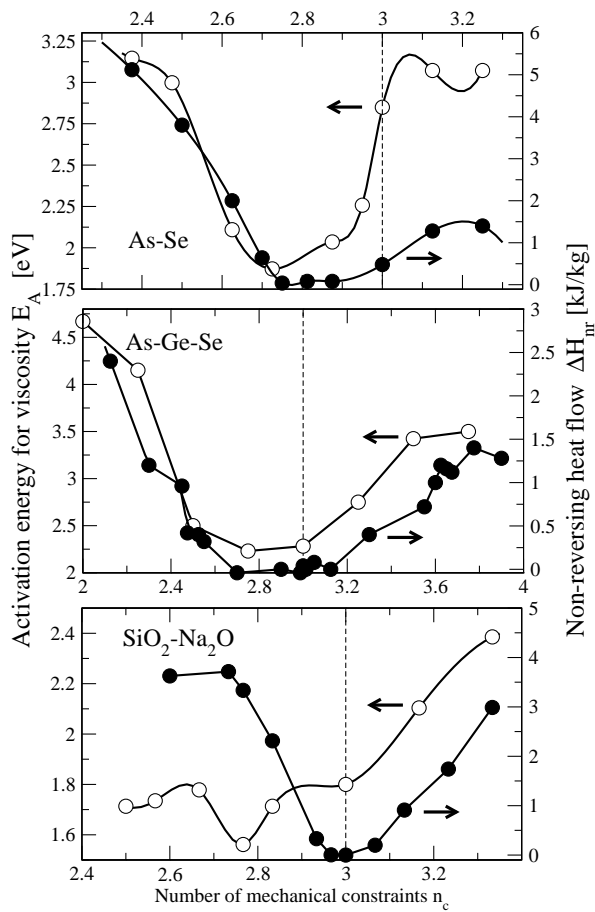}
\end{center}
\caption{\label{experim} Comparison between the activation energy for viscocity $E_A$ and the non-reversing heat flow $\Delta H_{nr}$ at the glass transition (right axis), as a function of the number of mechanical constraints $n_c$ for three selected glass systems: As-Se (data from \cite{AsSe1}, \cite{AsSe2}); As-Ge-Se (data from \cite{AsGeSe1}, \cite{AsGeSe2}; sodium silicates (data from \cite{nasi1}, \cite{nasi2}). The vertical broken lines indicate the isostatic composition.}
\end{figure}
This allows to define a fragility index $M$ from equ. (\ref{M}) which is represented in Figure \ref{area} (right axis) as a function of the number of mechanical constraints $n_c$. One can clearly remark that the fragility behaviour parallels the enthalpic changes at the glass transition. A minimum is found for both quantities when the number of floppy modes $f$ vanishes, i.e. when the glass is isostatic. This point corresponds to the location of the flexible to rigid transition \cite{Thorpe83} and the centroid of the intermediate phase \cite{Bool}.
\par
{\em Comparison with experiment.} Is there any experimental correlation between isostatic glasses and strong glass-forming liquids ? 
\par
Fairly complete viscosity and calorimetric measurements on several covalent glass-forming liquids are available in the literature. We focus on systems that undergo a flexible to rigid transition, i.e. oxide and chalcogenide glass in selected composition ranges that lead to a mean coordination number of $\bar r=2.4$ and $n_c=3$ \cite{JCP79},\cite{Thorpe83}. Compositional trends in the activation energy for viscosity $E_A$ for binary and ternary chalcogenide and oxide liquids appear in Fig. \ref{experim}. In the same figure (right axis) are also projected the non-reversing enthalpies $\Delta H_{nr}$ of corresponding glasses. The latter quantity provides an accurate measure of the enthalpic changes that have taken place during a heating/cooling cycle at the glass transition \cite{Bool,Bool1,AsSe2,AsGeSe1}. One can note that the global minima in the activation energy $E_A$ coincide at $n_c\simeq 3$ with those in $\Delta H_{nr}$ for chalcogenide glasses. Together with Fig. \ref{area}, these data demonstrate the correlation between the strong-fragile classification of glass forming liquids with the flexible-intermediate-stressed-rigid classification of corresponding glasses. The correlation unequivocally shows that Intermediate phase glasses where $n_c\simeq 3$ give rise to strong liquids, while both flexible ($n_c<3$) and stressed-rigid glasses ($n_c>3$) give rise to fragile liquids. 
\par
Chalcogenide glasses can be accurately described with a harmonic Keating potential \cite{Thorpe83,Mousseau}. Can the present conclusions be extended to glass-forming liquids characterized by potentials different from those shown in equ. (\ref{V}) ? The bottom panel of Figure \ref{experim} already sketches some limitations. In alkali silicate glasses, a large value for $n_c$ corresponds indeed to the silica-rich compositional region, i.e. to systems where the interaction can be fairly described by a Keating potential \cite{sio2_keat}. However, larger amounts of alkali ions (i.e. leading to lower $n_c$'s) increases the number of more weaker (Coulombic) interactions and should cancel the correlation. Weak alkali atom-non-bridging oxygen ionic bonds form, and as $T>T_g$, these weaker interactions cease to act as mechanical constraints enhancing the alkali-atoms mobilities and contributing to the fragility. In the modified oxides as in the H-bonded systems \cite{Phillips2008}, one therefore does not expect the glass-liquid correlation to uphold as in the chalcogenides.
\par
In summary, we have shown that a statistical model using a Keating potential with floppy modes was able to reproduce the generic features of the glass transition from a simple Monte Carlo dynamics. Glasses which are optimally constrained (isotatic) give rise to strong glass-forming liquids, and are found to display glass transitions with few enthalpic changes. Comparison with experiments shows that the demonstrated relationship holds for network glass-forming liquids. Limitations appear for systems where the nature of the potential $V$ is obviously changed in a more deeper fashion.
\par
It is a pleasure to acknowledge ongoing discussions with P. Boolchand,
B. Goodman, M. Malki and  P. Simon. LPTMC is  Unité Mixte de Recherche
du Centre National de la Recherche Scientifique (CNRS) n. 7600.

\end{document}